\documentstyle{article}
\font\cero=cmss10 scaled 1728
\font\uno=cmssbx10 scaled 1200
\begin{document}
\begin{flushleft}
{\cero On the symplectic structures for geometrical theories} \\[3em]
\end{flushleft} {\sf R. Cartas-Fuentevilla}\\
{\it Instituto de F\'{\i}sica, Universidad
Aut\'onoma de Puebla, Apartado postal J-48 72570 Puebla Pue., M\'exico*,
and Enrico Fermi Institute, University of Chicago, 5640 S. Ellis Ave.,
Chicago, Illinois 60637} \\ [4em]

We present a new approach for constructing covariant symplectic structures
for geometrical theories, based on the concept of adjoint operators. Such
geometric structures emerge by direct exterior derivation of underlying
symplectic potentials. Differences and similarities with other approaches
and future applications are discussed.

\noindent Running title:  On the symplectic....\\
\noindent * Permanent address

\begin{center}
{\uno I. INTRODUCTION}
\end{center}
\vspace{1em}

 Usually quantum field theories are studied by means of Feynman path
integrals or by means of canonical quantization. Path integral
quantization has the virtue to preserve all relevant symmetries,
including Poincar\'e invariance; however, the resultant theory has
not (unlike the canonical formalism) necessarily the standard
interpretation in terms of quantum mechanical states and
operators. On the other hand, the canonical formalism is
considered to be the antithesis of a manifestly covariant
treatment.

However, more recently, the essence of the canonical formulation
has been developed independently by  Witten {\it et al} \cite{1,2}
and Suckerman \cite{3} in such a way that manifestly preserves
Poincar\'e invariance as well as other relevant symmetries. Such a
formulation is based on a covariant description of Poisson
brackets in terms of a symplectic structure defined on the
manifold representing the phase space of classical solutions;
thus, quantization is carried out as the replacement of Poisson
brackets with commutators, and the resultant quantum theory will
be of the conventional type. Specifically, the Witten-{\it et al}
approach requires the construction, {\it a priori}, of a bilinear
product on variations of classical solutions. Subsequently, one
needs to verify that such a bilinear form corresponds to a
nondegenerate closed two-form on the phase space. Moreover, the
bilinear form must be a covariantly conserved current in its
spacetime dependence, as required for obtaining a symplectic
structure manifestly covariant. More specifically, in such a
description, the classical phase space is defined as {\it the
space of solutions of the classical equations of motion}; such a
definition is manifestly covariant. The construction of a
covariantly conserved two-form $J^{\mu}$ on such phase space
yields a symplectic structure $\omega$ defined as $\omega \equiv
\int_{\Sigma} J^{\mu} d \Sigma_{\mu}$ (being $\Sigma$ an initial
value hypersurface), independent of the choice of $\Sigma$ and, in
particular, Poincar\'e invariant. Additionally, in terms of the
symplectic structure $\omega$, the fact that Poisson brackets
satisfy the Jacoby identity, is equivalent that $\omega$ to be a
closed two-form on the phase space, which holds if $J^{\mu}$
itself is closed. With these properties, $J^{\mu}$ is known as
{\it the symplectic current}. Such a quantization scheme has been
applied, for example, for the analysis of two-dimensional gravity
(\cite{4} and references therein), and for the investigation of
the Wess--Zumino--Witten model on a circle \cite{5}.

Although in the present article we shall obtain essentially the
same geometric structures described above, the main novelty is
that such structures emerge in a direct and natural way using the
concept of adjoint operators. Particularly, the concept of
self-adjoint operators shows that, in the cases considered here,
there exist, in general, covariantly conserved currents, which
correspond on the phase space, to zero-, one-, and, two-forms.
Such differential forms are not independent, but that the two-form
turns out to be the exterior derivative of the corresponding
one-form, and correspond, thus, to an {\it exact} two-form (and
automatically to a closed two-form, as required for the symplectic
structure). In this manner, the present approach allows us to
find, unlike the Witten--Zuckerman procedure, fundamental
one-forms playing the role of {\it symplectic potentials} for the
theory.

In the next section, we shall discuss the key concept of adjoint
operators, and its consequences on the existence of covariantly conserved
currents. The interpretation of the terms involved in such a definition
as wedge products on the phase space is also discussed. In Section III
only the non-Abelian gauge theories and pure general relativity are
considered with the purpose of clarifying our basic ideas and to have a
direct comparison with previously known results, particularly with those
given in Reference [2]. In Section IV we shall finish with some concluding
remarks on our results and possible extensions of the present approach.

Concepts and definitions on differential forms, wedge products, exterior
derivative, etc., come entirely from Ref.\ \cite{2}. \\[2em]

\noindent {\uno II. ADJOINT OPERATORS AND CONSERVED CURRENTS}
\vspace{1em} \\

The general relationship between adjoint operators and covariantly
conserved currents has been already given in previous works
(\cite{7} and references cited therein), however, we shall discuss
it in this section for completeness.

If ${\cal P}$ is a linear partial differential operator that takes
matrix-valued tensor fields into themselves, then, the adjoint
operator of ${\cal P}$, is that operator ${\cal P}^{\dag}$, such
that
\begin{equation}
   {\rm Tr} \{ f^{\rho\sigma ...}[{\cal P}(g_{\mu\nu ...})]_{\rho\sigma ...}
   - [{\cal P}^{\dag}(f^{\rho\sigma ...})]^{\mu\nu ...}g_{\mu\nu ...} \} =
   \nabla_{\mu}J^{\mu},
\end{equation}
where $\rm Tr$ denotes the trace and $J^{\mu}$ is some vector field. From
this definition, if $\cal Q$ and $\cal R$ are any two linear operators,
one easily finds the following properties:
\[
   ({\cal Q}{\cal R})^{\dag} = {\cal R}^{\dag}{\cal Q}^{\dag}, \qquad
   ({\cal Q} + {\cal R})^{\dag} = {\cal Q}^{\dag} + {\cal R}^{\dag},
\]
and in the case of a function $F$,
\[
    F^{\dag} = F,
\]
which will be used implicitly below.

From Eq.\ (1) we can see that this definition automatically
guarantees that, if the field $f$ is a solution of the linear
system ${\cal P} (f) = 0$, and $g$ a solution of the adjoint
system ${\cal P}^{\dag} (g) = 0$, then we obtain the continuity
law $\nabla_{\mu}J^{\mu}=0$, which establishes that $J^{\mu}$ is a
covariantly conserved current depending on the fields $f$ and $g$.
This fact means that for any homogeneous equation system, one can
always construct a conserved current, taking into account the
adjoint system. This general result contains the self-adjoint case
(${\cal P}^{\dag} = {\cal P}$) as a particular one, for which $f$
and $g$ correspond to two independent solutions (in fact, the
cases treated in the present article are self-adjoint). Although
this result has been established assuming only tensor fields and
the presence of a single equation, such a result can be extended
in a direct way to equations involving spinor fields, matrix
fields, and the presence of more than one field \cite{6,7}.

Our main task in this work is to apply this very general result
for the analysis of the symplectic forms on the phase space of the
theories under consideration. Hence, it is important to clarify,
in the first instance, what the fields $f$, $g$, $J^{\mu}$ and the
differential operators ${\cal P}$, ${\cal P}^{\dag}$, and
$\nabla_{\mu}$ will mean on the phase space at the level of Eq.\
(1). First, such operators will depend only on the background
fields, and will correspond thus to zero-forms. Second, although
in our previous works we have identified the fields $f$ and $g$
with solutions of the equations governing the first-order
variations [${\cal P}(f) = 0 = {\cal P}(g)$], in the present work
we shall see that it is possible to find simultaneously that
${\cal P}(G) = 0$, where $G$ is some background field. Thus, since
the background fields and the first-order variations correspond,
on the phase space, to zero-forms and one-forms, respectively
\cite{2}, the left-hand side of Eq.\ (1) must be understood as a
wedge product, ${\rm Tr} \{ f \wedge {\cal P} (g) - {\cal
P}^{\dag}(f) \wedge g \}  = \nabla_{\mu} J^{\mu}$, on such phase
space, and something similar for the field $J^{\mu}$ in its
dependence on the fields $f$ and $g$. This subject will be clarified
in the examples below. \\[2em]

\noindent {\uno III. GEOMETRICAL THEORIES AND THEIR SYMPLECTIC STRUCTURES}
\vspace{1em} \\

In this section we shall see that the problem of finding the
symplectic structures (and in some cases the symplectic
potentials), is reduced to identify some fields satisfying some
homogeneous linear equations.

\vspace{1em}
\begin{center}
{\uno A. Yang--Mills theory}
\end{center}

Let us consider first the Yang--Mills equations:
\begin{equation}
     \partial\hspace{-.2cm}/^{\mu}F_{\mu\nu} = 0,
\end{equation}
where $F_{\mu\nu} = \partial \hspace{-.2cm}/_{\mu} A_{\nu} - \partial
\hspace{-.2cm}/_{\nu}A_{\mu}$ is the Yang-Mills curvature, $A_{\mu}$ the
background gauge connection, and $\partial \hspace{-.2cm}/_{\mu} \equiv
\partial_{\mu} + [A_{\mu},\hspace{.3cm}]$, the gauge covariant derivative.

From Eq.\ (2), the variations of the background fields are governed by the
equations:
\begin{equation}
     \partial\hspace{-.2cm}/^{\mu} \delta F_{\mu\nu} + [\delta A^{\mu},
F_{\mu\nu}] = [\delta^{\alpha}_{\nu} \partial \hspace{-.2cm}/^{\mu}
\partial \hspace{-.2cm} /_{\mu} - \partial \hspace{-.2cm}/^{\alpha}
\partial \hspace{-.2cm} /_{\nu}] \delta A_{\alpha} + [\delta A_{\alpha},
{F^{\alpha}}_{\nu}] \equiv [{\cal P} (\delta A_{\alpha})]_{\nu} = 0,
\end{equation}
where $\delta F_{\mu\nu} = \partial \hspace{-.2cm}/_{\mu} \delta
A_{\nu} - \partial \hspace{-.2cm}/_{\nu} \delta A_{\mu}$, is the
variation of the curvature, $\delta A_{\mu}$ the variation of the
gauge connection, and the operator ${\cal P}$ is a homogeneous
linear operator depending only on the background fields. Up to
here, the usual equations for the Yang--Mills fields and their
variations.

Now the idea is to apply our present approach for obtaining all on
the symplectic structure for the theory, directly from the basic
equations (2) and (3). For this purpose, let $B_{\mu}$ and
$C_{\mu}$ be any two matrix-valued fields (which will be
identified below as a pair of gauge connection variations in one
case, and as the background gauge connection and its variation in
the  particular case of Abelian fields), and using the explicit
form of the operator ${\cal P}$ in Eq.\ (3), we have that
\begin{equation}
   B^{\nu} \wedge[{\cal P} (C_{\alpha})]_{\nu} - [{\cal P}
   (B^{\nu})]^{\alpha} \wedge C_{\alpha} = \partial \hspace{-.2cm}/_{\mu}
   [B^{\nu} \wedge (\partial \hspace{-.2cm}/^{\mu} C_{\nu} - \partial
   \hspace{-.2cm}/_{\nu} C^{\mu}) + (\partial \hspace{-.2cm}/^{\nu}
   B^{\mu} - \partial \hspace{-.2cm}/^{\mu} B^{\nu}) \wedge C_{\nu}] +
   [F_{\mu\nu},B^{\mu} \wedge C^{\nu}],
\end{equation}
where
\begin{equation}
  B^{\nu} \wedge [C_{\alpha},{F^{\alpha}}_{\nu}] - [B_{\alpha},
  {F^{\alpha}}_{\nu}] \wedge C^{\nu} = [F_{\mu\nu}, B^{\mu} \wedge
  C^{\nu}],
\end{equation}
and identities of the form $B^{\nu} \wedge \partial
\hspace{-.2cm}/^{\alpha} \partial \hspace{-.2cm}/_{\nu} C_{\alpha}
\equiv
\partial \hspace{-.2cm}/_{\mu} (B^{\nu} \partial
\hspace{-.2cm}/_{\nu} C^{\mu} - \partial \hspace{-.2cm}/^{\nu}
B^{\mu} \wedge C_{\nu}) + \partial\hspace{-.2cm}/_{\nu} \partial
\hspace{-.2cm}/^{\alpha} B^{\nu} \wedge C_{\alpha}$ have been
used. Taking the trace of Eq.\ (4), we obtain
\begin{equation}
    {\rm Tr}[ B^{\nu} \wedge[{\cal P} (C_{\alpha})]_{\nu} - [{\cal P}
(B^{\nu})]^{\alpha} \wedge C_{\alpha} ] = \partial_{\mu} {\rm Tr} [B_{\nu}
\wedge \partial \hspace{-.2cm}/^{[\mu} C^{\nu]} - \partial
\hspace{-.2cm}/^{[\mu} B^{\nu]} \wedge C_{\nu}]
\end{equation}
which has the form of Eq.\ (1) with ${\cal P} = {\cal P}^{\dag}$. Thus, we
can obtain the continuity equation:
\begin{equation}
   \partial_{\mu}J^{\mu}=0, \qquad J^{\mu} \equiv {\rm Tr} [B_{\nu} \wedge
\partial \hspace{-.2cm}/^{[\mu} C^{\nu]} - \partial \hspace{-.2cm}/^{[\mu}
B^{\nu]} \wedge C_{\nu}],
\end{equation}
provided that
\begin{equation}
   [{\cal P}(C_{\alpha})]_{\nu} = 0, \quad {\rm and} \quad [{\cal P}
(B^{\nu})]^{\alpha} = 0.
\end{equation}

As we shall see, the whole physical information about the
covariant symplectic structure of the Yang--Mills theory is
contained in Eq.\ (7), it remains only to identify the fields
$B_{\mu}$ and $C_{\mu}$ satisfying Eqs.\ (8). In according to Eq.\
(3), the obvious case is to choose such fields as a pair of
variations, say $B_{\mu}=\delta A^{1}_{\mu}$, and $C_{\mu}=\delta
A^{2}_{\mu}$ (they have not to correspond necessarily to the same
variation). In this manner, $J^{\mu}$ in Eq.\ (7) corresponds, in
this case, to the following (nondegenerate) two-form on the phase
space:
\begin{equation}
  J_{\mu} = {\rm Tr}[ \delta A^{\nu}_{1} \wedge \delta F^{2}_{\mu\nu}-
\delta F^{1}_{\mu\nu}\wedge\delta A^{\nu}_{2} ] \nonumber \\
  = \frac{1}{2} \delta{\rm Tr}[ A^{\nu}_{1} \delta F^{2}_{\mu\nu} -
F^{2}_{\mu\nu} \delta A^{\nu}_{1}- F^{1}_{\mu\nu} \delta A^{\nu}_{2}
+ A^{\nu}_{2} \delta F^{1}_{\mu\nu} ] \nonumber \\
\equiv \delta \theta_{\mu},
\end{equation}
where $F^{\rm i}_{\mu\nu} = \partial \hspace{-.2cm}/_{\mu}
A^{\rm i}_{\nu} - \partial \hspace{-.2cm}/_{\nu} A^{\rm i}_{\mu}$, $\delta
F^{\rm i}_{\mu\nu} = \partial \hspace{-.2cm}/_{\mu} \delta A^{\rm i}_{\nu}
- \partial \hspace{-.2cm}/_{\nu}\delta A^{\rm i}_{\mu}$ $({\rm i} = 1,2)$,
and we have used the Leibniz rule for the exterior derivative $\delta$,
and the fact that $\delta^{2} = 0$. In particular, if $\delta A^{1}_{\mu}
= \delta A_{\mu} = \delta A^{2}_{\mu}$, from Eq.\ (9) $J_{\mu} = 2{\rm Tr}
(\delta A^{\nu} \wedge \delta F_{\mu\nu})$, which is essentially the
Crncovi\'c-Witten current \cite{2}. Furthermore, we have defined the
one-form $\theta_{\mu}$ as
\begin{equation}
    \theta_{\mu} \equiv \frac{1}{2}{\rm Tr}[ A^{\nu}_{1} \delta
F^{2}_{\mu\nu} - F^{2}_{\mu\nu} \delta A^{\nu}_{1} - F^{1}_{\mu\nu} \delta
A^{\nu}_{2} + A^{\nu}_{2} \delta F^{1}_{\mu\nu} ],
\end{equation}
in this manner, $\theta_{\mu}$ is the {\it symplectic potential}
for the theory. Note that, according to Eq.\ (9), the symplectic
potential is defined up to the exterior derivative of any
matrix-valued field $\lambda_{\mu}$: $J_{\mu} = \delta
(\theta_{\mu} + \delta \lambda_{\mu})$.

In the particular case of Abelian fields, from Eqs.\ (2) and (3)
we have that $[{\cal P}(A_{\alpha})]_{\nu} = 0$, where
$A_{\alpha}$ is the background gauge connection. In this manner,
we can identify $B_{\nu} = \delta A^{1}_{\nu}$ and $C_{\nu}=
A^{2}_{\nu}$ (a variation and a background gauge connection
respectively), and then the symplectic potential $\theta_{\mu}$
given in Eq.\ (10) is [like the corresponding symplectic current
in Eq.\ (9)] covariantly conserved. Moreover, we can identify for
Abelian fields $B_{\nu} = A^{1}_{\nu}$, and $C_{\nu} =
A^{2}_{\nu}$ (a pair of background fields) in Eq.\ (8); thus, from
Eq.\ (7) $J_{\mu} = A^{\nu}_{1} F^{2}_{\mu\nu} - A^{\nu}_{2}
F^{1}_{\mu\nu}$, which is a covariantly conserved zero-form on the
phase space (a conserved current for the exact theory).

Since $J_{\mu}$ in Eq.\ (9) is an {\it exact} two-form (it comes
from the variations of the symplectic potential $\theta_{\mu}$),
corresponds automatically to a closed two-form ($\delta J^{\mu} =
\delta^{2} \theta^{\mu} = 0$), as required for the symplectic
structure. Unlike the Crncovi\'c--Witten approach, we do not need
to verify the covariant conservation of our symplectic current,
such a property is guaranteed for Eqs.\ (7) and (8). Therefore,
$\omega = \int_{\Sigma} J^{\mu} d \Sigma_{\mu}$ is the symplectic
structure with the wanted properties for the Yang--Mills theory
\cite{2}. Moreover, since the present symplectic structure is
essentially the Crncovi\'c-Witten result, has the same invariance
properties under gauge transformations \cite{2}. Specifically, as
shown in Ref.\ \cite{2}, under gauge transformations of the gauge
connection $A_{\mu}\rightarrow A_{\mu} +
\partial_{\mu}\varepsilon + [A_{\mu},\varepsilon]$, $\delta A^{\rm i}_{\mu}$ and
$\delta F^{\rm i}_{\mu\nu}$ transform homogeneously, and then
$J^{\mu}$ and $\omega$ are gauge invariant. Furthermore, following
Ref.\ \cite{2}, one can verify easily that $\omega$ has vanishing
components in the gauge directions in field space [see Eq.\ (30)
in such Reference], which allows us to construct the symplectic
structure on the corresponding gauge-invariant space (reduced
phase space).

The above results are obtained displaying explicitly the variation
of the gauge connection $\delta A_{\alpha}$ in Eq.\ (3). However,
it is not the only way for obtaining such results. One can
consider Eq.\ (3) in its original form $\partial
\hspace{-.2cm}/^{\mu} \delta F_{\mu\nu} + [\delta A^{\mu},
F_{\mu\nu}] = 0$, and the relation $\delta F_{\mu\nu} =
\partial \hspace{-.2cm}/_{\mu} \delta A_{\nu} - \partial
\hspace{-.2cm}/_{\nu} \delta A_{\mu}$, as a system of equations
governing the field variations $\delta F_{\mu\nu}$, and $\delta
A_{\mu}$, considering them as independent field variables:
\[
   \left[ \begin{array}{cc} \partial \hspace{-.2cm}/^{\mu} &
- [{F^{\alpha}}_{\nu}, \hspace{.3cm}] \\
   1 & (\delta^{\alpha}_{\mu} \partial \hspace{-.2cm}/_{\nu} -
\delta^{\alpha}_{\nu} \partial \hspace{-.2cm}/_{\mu} ) \end{array} \right]
\left[ \begin{array}{c} \delta F_{\mu\nu} \\
   \delta A_{\alpha} \end{array} \right] = 0, \nonumber
\]
and using again the definition (1) with ${\cal P}$ now being the
matrix operator in the preceding equation, one obtains essentially
the same results.

\vspace{1em}
\begin{center}
{\uno B. General relativity}
\end{center}

The variations of the vacuum Einstein equations $R_{\mu\nu}=0$ are
\begin{equation}
    \nabla_{\alpha} \delta \Gamma^{\alpha}_{\mu\nu} - \nabla_{\mu} \delta
\Gamma^{\alpha}_{\nu\alpha} = 0,
\end{equation}
where $\nabla_{\alpha}$ is the covariant derivative compatible
with the background metric $g_{\mu\nu}$, and $\delta
\Gamma^{\alpha}_{\mu\nu} = \frac{1}{2} g^{\alpha\beta}
(\nabla_{\mu} \delta g_{\nu\beta} + \nabla_{\nu} \delta
g_{\mu\beta} - \nabla_{\beta} \delta g_{\mu\nu})$, the variation
of the metric connection \cite{2}. Displaying explicitly the
metric variations $\delta g_{\mu\nu}$, Eqs.\ (11) take the form
\begin{equation}
    [g^{\alpha}_{\nu} \nabla^{\beta} \nabla_{\mu} + g^{\alpha}_{\mu}
\nabla^{\beta} \nabla_{\nu} - g^{\alpha}_{\mu} g^{\beta}_{\nu}
\nabla^{\rho} \nabla_{\rho} - g^{\alpha\beta} \nabla_{\mu} \nabla_{\nu} +
g_{\mu\nu} (g^{\alpha\beta} \nabla^{\rho} \nabla_{\rho} - \nabla^{\beta}
\nabla^{\alpha})] \delta g_{\alpha\beta} = 0,
\end{equation}
which can be written in a compact form as
\begin{equation}
    [{\cal E} (\delta g_{\alpha\beta})]_{\mu\nu}=0,
\end{equation}
where ${\cal E}$ is the linear operator (depending only on the
background fields) appearing in Eq.\ (12). With the same idea of
the above case, let $A_{\mu\nu}$, and $B_{\mu\nu}$ be any two
2-index (symmetric) tensor fields (in the first case these fields
will be identified as a pair of metric variations for
constructing the symplectic current, and as the background metric
and a metric variation in the second case for obtaining the
corresponding symplectic potential), and using the explicit form
of the operator ${\cal E}$,we have that
\begin{equation}
    B^{\mu\nu} \wedge [{\cal E} (A_{\alpha\beta})]_{\mu\nu} -
[{\cal E}(B^{\mu\nu})]_{\alpha\beta} \wedge A^{\alpha\beta} =
\nabla_{\mu} S^{\mu\alpha\beta\lambda\rho\gamma} (B_{\alpha\beta} \wedge
\nabla_{\lambda} A_{\rho\gamma} -\nabla_{\lambda} B_{\rho\gamma} \wedge
A_{\alpha\beta}),
\end{equation}
where
\begin{equation}
    S^{\mu\alpha\beta\lambda\rho\gamma} = g^{\mu(\rho} g^{\gamma) (\alpha}
g^{\beta)\lambda} - \frac{1}{2} g^{\mu\lambda} g^{\alpha(\rho}
g^{\gamma)\beta} - \frac{1}{2} g^{\mu(\alpha} g^{\beta)\lambda}
g^{\rho\gamma} - \frac{1}{2} g^{\alpha\beta} g^{\mu(\rho}
g^{\gamma)\lambda} + \frac{1}{2} g^{\alpha\beta} g^{\mu\lambda}
g^{\rho\gamma}.
\end{equation}
Like the Yang--Mills case, Eq.\ (14) has the form of Eq.\ (1) with
${\cal E} = {\cal E}^{\dag}$. Then, we obtain the local continuity
equation:
\begin{equation}
 \nabla_{\mu}J^{\mu} = 0, \qquad J^{\mu} \equiv
S^{\mu\alpha\beta\lambda\rho\gamma} (B_{\alpha\beta} \wedge
\nabla_{\lambda} A_{\rho\gamma} - \nabla_{\lambda} B_{\rho\gamma} \wedge
A_{\alpha\beta}),
\end{equation}
provided that
\begin{equation}
    [{\cal E} (A_{\alpha\beta})]_{\mu\nu} = 0, \quad {\rm and} \quad
[{\cal E} (B^{\mu\nu})]_{\alpha\beta} = 0.
\end{equation}
In according to Eq.\ (13), an obvious identification for the
fields $A_{\mu\nu}$, and $B_{\mu\nu}$ satisfying Eqs.\ (17) is
\begin{equation}
   A_{\alpha\beta} = \delta g^{1}_{\alpha\beta}, \quad {\rm and} \quad
B_{\mu\nu} = \delta g^{2}_{\mu\nu},
\end{equation}
we mean, a pair of variations. In this manner, from Eq.\ (16),
\begin{equation}
   J^{\mu}=S^{\mu\alpha\beta\lambda\rho\gamma} (\delta g^{2}_{\alpha\beta}
\wedge \nabla_{\lambda} \delta g^{1}_{\rho\gamma} - \nabla_{\lambda}
\delta g^{2}_{\rho\gamma} \wedge \delta g^{1}_{\alpha\beta}),
\end{equation}
corresponds to a covariantly conserved two-form on the phase space. The
last expression can be rewritten, using Eq.\ (15), in terms of the
variations of the metric connection:
\begin{equation}
     J^{\mu} = (\delta \Gamma^{\mu}_{\alpha\beta})_{1} \wedge \left[
\delta g^{\alpha\beta}_{2} + \frac{1}{2} g^{\alpha\beta} (\delta \ln
g)_{2} \right] - (\delta \Gamma^{\nu}_{\alpha\nu})_{1} \wedge \left[
\delta g^{\mu\alpha}_{2} + \frac{1}{2} g^{\mu\alpha} (\delta \ln g)_{2}
\right] - (1 \leftrightarrow 2),
\end{equation}
where $(\delta \Gamma^{\mu}_{\alpha\beta})_{1} = \frac{1}{2}
g^{\mu\rho} \left[ \nabla_{\alpha} \delta g^{1}_{\beta\rho} +
\nabla_{\beta} \delta g^{1}_{\alpha\rho} - \nabla_{\rho} \delta
g^{1}_{\alpha\beta} \right]$, $(\delta \ln g)_{2} = g^{\mu\nu}
\delta g^{2}_{\mu\nu} = - g_{\mu\nu} \delta g^{\mu\nu}_{2}$, and
$(1 \leftrightarrow 2)$ means a term similar to the first one,
just interchanging the subscripts 1 and 2, such as Eq.\ (19). If
we set $\delta g^{1}_{\mu\nu} = \delta g^{2}_{\mu\nu} = \delta
g_{\mu\nu}$, $J^{\mu}$ in Eq.\ (20) reduces exactly to the
Crncovi\'c--Witten current [see Eq.\ (34) of Ref.\ \cite{2}].

However, the choice (18) for the fields $A_{\alpha\beta}$, and
$B_{\mu\nu}$, is not the unique one for satisfying Eqs.\ (17). We
can keep $A_{\alpha\beta} = \delta g^{1}_{\alpha\beta}$, but to
identify $B_{\mu\nu}$ as the background metric $g_{\mu\nu}$, since
$\nabla_{\lambda} g_{\mu\nu} = 0$, and the explicit form of ${\cal
E}$ in Eq.\ (12), we have that
\begin{equation}
    [{\cal E}(g_{\mu\nu})]_{\alpha\beta} = 0.
\end{equation}
Therefore, from Eq.\ (16), we have that the one-form
\begin{equation}
   \theta^{\mu} \equiv S^{\mu\alpha\beta\lambda\rho\gamma}
g_{\alpha\beta} \nabla_{\lambda} \delta g^{1}_{\rho\gamma},
\end{equation}
is also a covariantly conserved current on the phase space.
$\theta^{\mu}$ can also be rewritten in terms of the variations of
the metric connection:
\begin{equation}
     \theta^{\mu} = g^{\mu\alpha} (\delta \Gamma^{\nu}_{\alpha\nu})_{1} -
g^{\alpha\beta}  (\delta \Gamma^{\mu}_{\alpha\beta})_{1}.
\end{equation}

Moreover, the conserved currents $J^{\mu}$ and $\theta^{\mu}$ given in
Eqs.\ (19)--(20) and (22)--(23) respectively, are not independent.
Considering that $\delta \sqrt{g} = \frac{1}{2} \sqrt{g} \delta \ln g$,
from Eq.\ (23), we have that
\begin{eqnarray}
    \delta (\sqrt{g} \theta^{\mu}) \!\! & = & \!\! \sqrt{g} \left[ \delta
g^{\mu\alpha}_{2} \wedge (\delta \Gamma^{\nu}_{\alpha\nu})_{1} - \delta
g^{\alpha\beta}_{2} \wedge (\delta \Gamma^{\mu}_{\alpha\beta})_{1} \right]
\nonumber \\
     \!\! &  & \!\! - \frac{1}{2} \sqrt{g} \left[ g^{\mu\alpha} (\delta
\Gamma^{\nu}_{\alpha\nu})_{1} - g^{\alpha\beta} (\delta
\Gamma^{\mu}_{\alpha\beta})_{1} \right] \wedge (\delta \ln g)_{2},
\end{eqnarray}
where we have considered also that $ \delta^{2} = 0$, the Leibniz rule for
the exterior derivative, and a variation of the background metric
appearing in Eq.\ (23) in general different of $\delta g^{1}_{\mu\nu}$,
and denoted conveniently by $\delta g^{2}_{\mu\nu}$. Making a direct
comparison, the right-hand side of Eq.\ (24) corresponds, by a factor of
$\sqrt{g}$, to the first term on the right-hand side of Eq.\ (20). With an
interchange of the superscripts 1 and 2 in Eq.\ (24) (which corresponds to
identify $A_{\alpha\beta}$ with the background metric and $B_{\mu\nu}$ with
the metric variation),  we obtain essentially the second term on the
right-hand side of Eq.\ (20). In this manner, we can rewrite
\begin{equation}
    \theta^{\mu} = S^{\mu\alpha\beta\lambda\rho\gamma} (g^{2}_{\alpha\beta}
\nabla_{\lambda} \delta g^{1}_{\rho\gamma} + g^{1}_{\alpha\beta}
\nabla_{\lambda} \delta g^{2}_{\rho\gamma}),
\end{equation}
and then,
\begin{equation}
    \delta (\sqrt{g} \theta^{\mu}) = \sqrt{g} J^{\mu},
\end{equation}
which means that $\sqrt{g} J^{\mu}$ is an {\it exact} two-form, and
$\sqrt{g} \theta^{\mu}$ is then the {\it symplectic potential} for the
theory (which is defined up to the exterior derivative of any vector
field). Since $\nabla_{\lambda} g_{\mu\nu} = 0$ and $g = g(g_{\mu\nu})$,
$\sqrt{g} \theta^{\mu}$ and $\sqrt{g} J^{\mu}$ are, like $\theta^{\mu}$
and $J^{\mu}$, also covariantly conserved.

In the Crncovi\'c-Witten approach, one needs to show that $\nabla_{\mu}
J^{\mu} = 0$; in the present approach $J^{\mu}$ comes directly from the
continuity equation (16).  Moreover, from Eq.\ (26), $\sqrt{g} J^{\mu}$ is
an exact two-form, and automatically a closed two-form, as required for the
symplectic structure $\omega = \int_{\Sigma} \sqrt{g} J^{\mu} d
\Sigma_{\mu}$, which has the wanted properties. Since $\omega$ is
essentially that given in Ref.\ [2], has the same invariance properties
under gauge transformations described in such reference.

If we choice $A_{\mu\nu}=g^{1}_{\mu\nu}$, and $B_{\alpha\beta} =
g^{2}_{\alpha\beta}$ (a pair of background solutions), both satisfying Eq.\
(21), then from  the local equation (16), we have that $J^{\mu}=0$, which
means that there no exist a (local) conserved current for the exact theory
different to the trivial one.

Finally, if we consider Eq.\ (11) and the relation between $\delta\Gamma$
and $\delta g_{\mu\nu}$ as a system for these field variations
(considering them as independent), one obtains essentially the same
results. \\[2em]

\begin{center}
{\uno IV. CONCLUDING REMARKS}
\end{center}
\vspace{1cm}

As we have seen, the present approach based on the concept of
(self-)adjoint operators leads, in a rigorous way, to local
continuity laws for the theory under study. Such continuity
equations disclose the existence of different conserved currents,
in particular those associated with a covariant description of the
corresponding symplectic structure.

The symplectic structures described in Ref.\ [1-3], are always
related to a pair of solutions of the equations governing the
variations of classical solutions. In the present scheme, the
self-adjoint case corresponds, as we have seen in the examples, to
that case. Nevertheless, as discussed  in Sec.\ II, there exists a
more general case, which establishes the possibility of
constructing a (nondegenerate) two-form related to a solution of
the equations governing the variations, and a solution of the
corresponding adjoint system. No such possibility was previously
known in the literature. However, such a two-form is not
necessarily closed, remaining to study under what conditions this
two-form represents a symplectic structure. In fact, there are
several cases in physics involving operators that are not
self-adjoint, where the present approach will be useful: usual
free massless fields equations of spin greater that one on a
curved spacetime, equations for first-order variations coming from
string-inspired actions, etc. Works along these lines are in
progress and will be the subject of forthcoming communications.

On the other hand, the Zuckerman formalism, unlike the present
one, requires an explicit extension for covering fermionic fields
\cite{3}. Even though in the present article we have limited our
discussion to bosonic field theories, the adjoint operator
formalism allows us to treat bosonic and fermionic fields (and
the simultaneous presence of both) on the same footing, since the
fundamental definition (1), which is our starting point, extends
for spinor fields \cite{6}. In this case, we are particularly
interested in superstring theory, and works along these lines are
also in progress.

Finally, the connection between adjoint operators and conserved
currents used in the present article, has been also used in Ref.
[7], although for a different purpose: for obtaining conserved
quantities from non-Hermitian systems. In this manner, a scheme
based on adjoint operators has different ramifications of wide
interest in physics, whose applications also will be
the aim of future investigations. \\[2em]

\begin{center}
{\uno ACKNOWLEDGMENT}
\end{center}
\vspace{1em}

This work was supported by CONACyT and the Sistema Nacional de
Investigadores (M\'exico). The author wants to thank Professor Robert Wald
for the kind hospitality provided at the Enrico Fermi Institute,
University of Chicago. \\[2em]

\begin{center}
{\uno  APPENDIX: NO `PUZZLE' FOR THE SYMPLECTIC CURRENT}
\end{center}
\renewcommand{\theequation}{A\arabic{equation}}
\setcounter{equation}{0}
\vspace{1em}

In the Crncovi\'c--Witten approach \cite{2}, unlike the present
scheme, there is not a procedure for ob\-tai\-ning the explicit
form for the symplectic structure (or for the symplectic
potential). In fact, it may be very difficult to guess such an
explicit form for more general and complicated cases. However,
without invoking the concept of adjoint operators used in the
present scheme, one may be able of obtaining the explicit form of
the potential symplectic, starting directly from the basic
equations for the variations. For example, in the general
relativity case, the equation (11) for the variations can be
rewritten in the form
\begin{equation}
    \nabla_{\alpha} (\delta \Gamma^{\alpha}_{\mu\nu} -
\delta^{\alpha}_{\mu} \delta \Gamma^{\lambda}_{\nu\lambda}) = 0,
\end{equation}
which implies that the tensor field $T^{\alpha}_{\mu\nu} \equiv
\delta \Gamma^{\alpha}_{\mu\nu} - \delta^{\alpha}_{\mu} \delta
\Gamma^{\lambda}_{\nu\lambda}$ is a covariantly conserved one-form
on the phase space. Since $\nabla_{\lambda} g_{\mu\nu} = 0$, and
$g = g(g_{\mu\nu})$, the one-form $T^{\alpha} \equiv \sqrt{g}
g^{\mu\nu} T^{\alpha}_{\mu\nu}$ is also covariantly conserved:
$\nabla_{\alpha} T^{\alpha} = 0$. Using Eq.\ (23), it is very easy
to find that $g^{\mu\nu} T_{\mu\nu}^{\alpha} = \theta^{\alpha}$,
thus $T^{\alpha} = \sqrt{g} \theta^{\alpha}$. In this manner,
$T^{\alpha}$ coming from Eq.\ (A1), is the symplectic potential,
whose variations gene\-ra\-te automatically a closed two-form.
However, regardless of adjoint operators, one must verify the
covariant conservation of such a two-form in order to obtain a
covariant
description. \\[2em]

\end{document}